\documentclass[12pt]{article}

\usepackage{latexsym}
\def\thefootnote{\fnsymbol{footnote}}
\newcommand{\myref}[1]{(\ref{#1})}
\def\tF{{\tilde F}}

\def\Del{\Delta}
\def\half{{1\over2}}
\def\bea{\begin{eqnarray}}
\def\eea{\end{eqnarray}}
\def\beq{\begin{equation}}
\def\eeq{\end{equation}}
\def\bL{\bar{\Lambda}}
\def\ux{$U(1)_X$}
\def\uk{$U(1)_K$}
\def\uo{$U(1)$}
\def\dx{\delta_X}
\def\vx{V_X}
\def\superint{\int d^{4}\theta}

\newcommand{\Da}{{\cal D}_{\alpha}}

\newcommand{\Dd}{{\cal D}_{\dot{\beta}}}
\newcommand{\Wa}{W_{\alpha}}

\newcommand{\Wc}{W^{\alpha}}

\newcommand{\Xa}{X_{\alpha}}
\newcommand{\Xc}{X^{\alpha}}

\def\im{{\rm Im}}

\def\D{{\cal D}}
\def\bD{\bar{\D}}
\def\pp{\partial}
\def\ibar{\bar{\imath}}
\def\bj{\bar{\jmath}}
\def\[{\left [}
\def\]{\right ]}
\def\({\left (}
\def\){\right )}
\def\lbr{\left\{}
\def\rbr{\right\}}
\def\r{\right|}
\def\l{\left.}
\def\bPh{\bar{\Phi}}
\def\H{\bar{H}}
\def\Z{{\bar{Z}}}
\def\T{\bar{T}}
\def\z{\bar{z}}
\def\S{{\bar{S}}}
\def\Tr{{\rm Tr}}
\def\bR{\bar{R}}
\def\L{{\cal L}}

\def\cM{{\cal{M}}}

\def\chiproj{(\bD^2 - 8R)}
\def\bchiproj{(\D^2 - 8 \bar R)}

\def\ddd{\nonumber \\ &&}

\def\hc{ + {\rm h.c.}}

\begin{document}

\begin{titlepage}
\begin{center}

\hfill LBNL-2032E \\
\hfill UCB-PTH-09/19 \\
\hfill NSF-KITP-09-106\\
\hfill arXiv:0906.3505 [hep-th] \\
\hfill Aug 2009 \\[.3in]

{\large {\bf Anomaly structure of supergravity and anomaly 
cancellation}}\footnote{This work was supported in part
by the Director, Office of Science, Office of High Energy and Nuclear
Physics, Division of High Energy Physics, of the U.S. Department of
Energy under Contract DE-AC02-05CH11231, in part by the National
Science Foundation under grants PHY-0457315 and PHY99-07949.}  \\[.2in]

Daniel Butter {\em and} Mary K. Gaillard
\\[.1in]

{\em Department of Physics and Theoretical Physics Group,
 Lawrence Berkeley Laboratory, 
 University of California, Berkeley, California 94720}\\[.5in] 

\end{center}

\begin{abstract}
We display the full anomaly structure of supergravity, including new
contributions to the conformal anomaly.  Our result has
the super-Weyl and K\"ahler \uo\, transformation properties that are
required for implementation of the Green-Schwarz mechanism for
anomaly cancellation.

\end{abstract}
\end{titlepage}

\newpage

\renewcommand{\thepage}{\roman{page}}
\setcounter{page}{2}
\mbox{ }

\vskip 1in

\begin{center}
{\bf Disclaimer}
\end{center}

\vskip .2in

\begin{scriptsize}
\begin{quotation}
This document was prepared as an account of work sponsored by the United
States Government. While this document is believed to contain correct 
 information, neither the United States Government nor any agency
thereof, nor The Regents of the University of California, nor any of their
employees, makes any warranty, express or implied, or assumes any legal
liability or responsibility for the accuracy, completeness, or usefulness
of any information, apparatus, product, or process disclosed, or represents
that its use would not infringe privately owned rights.  Reference herein
to any specific commercial products process, or service by its trade name,
trademark, manufacturer, or otherwise, does not necessarily constitute or
imply its endorsement, recommendation, or favoring by the United States
Government or any agency thereof, or The Regents of the University of
California.  The views and opinions of authors expressed herein do not
necessarily state or reflect those of the United States Government or any
agency thereof, or The Regents of the University of California.
\end{quotation}
\end{scriptsize}

\vskip 2in

\begin{center}
\begin{small}
{\it Lawrence Berkeley Laboratory is an equal opportunity employer.}
\end{small}
\end{center}

\newpage
\renewcommand{\theequation}{\arabic{equation}}
\renewcommand{\thepage}{\arabic{page}}
\setcounter{page}{1}
\def\thefootnote{\arabic{footnote}}
\setcounter{footnote}{0}

When compactified from ten to four space-time dimensions, the weakly
coupled heterotic string theory~\cite{wchs} has an invariance under a
discrete group of transformations known as ``T-duality'' or ``target
space modular invariance''~\cite{mod}.  The effective four dimensional
(4d) theory includes several important ``moduli'' chiral
supermultiplets: the dilaton supermultiplet $S$, whose vacuum value
determines the gauge coupling constant and the $\theta$-parameter of
the 4d gauge theory, and ``K\"ahler moduli'' $T^i$ whose vacuum values
determine the radii of compactification.  The T-duality invariance of
the effective 4d supergravity theory results in several desirable
features~\cite{gnrev}: 1) it assures that the K\"ahler moduli, or
``T-moduli'' are generically stabilized at self-dual points, with
vanishing vacuum values for their auxiliary fields, so that supersymmetry
breaking is dilaton dominated and no large flavor mixing is induced;
2) it protects a symmetry known as ``R-symmetry'' that assures that
the mass of the axion (pseudoscalar) component of the dilaton
supermultiplet remains sufficiently small to offer a solution to the
strong CP problem; and 3) it may provide a residual discrete symmetry
at low energy that plays the role of R-parity, needed to
preserve lepton and baryon number conservation and the stability of
the lightest supersymmetric partner, which makes the latter an
attractive candidate for dark matter.  This symmetry can be stronger
than R-parity and thus forbid higher dimension operators that could
otherwise generate too large an amplitude for proton decay.

At the quantum level of the effective theory, T-duality is broken by
quantum anomalies, as is, generically, an Abelian \ux\, gauge
symmetry, both of which are exact symmetries of string perturbation
theory.  It was realized some time ago that these symmetries could be
restored by a combination of 4-d counterparts~\cite{gs4} of the
Green-Schwarz (GS) mechanism in 10 dimensions~\cite{gs} and string
threshold corrections~\cite{th}.  However anomaly cancellation has
been demonstrated explicitly only for the coefficient of the
Yang-Mills superfield strength bilinear.  The entire supergravity
chiral anomaly has in fact been determined~\cite{danf}, but the
complete superfield form of the anomaly is required to fully implement
anomaly cancellation.

The anomaly arises from linear and logarithmic divergences in the
effective supergravity theory, and is ill-defined in an
unregulated theory.  We use Pauli Villars (PV) regulation,
which has been shown~\cite{pv} to require only massive chiral multiplets
and Abelian gauge multiplets as PV regulator fields, thereby preserving,
for example, BRST invariance. 

T-duality acts as follows on chiral (antichiral)
superfields $Z^p = T^i,\Phi^a\;(\Z^{\bar p} = \T^{\ibar},\bPh^{\bar a})$:
\beq T^i\to h(T^j), \qquad \Phi^a \to
f(q^a_i,T^j)\Phi^a,\qquad \T^{\ibar}\to h^*(\T^{\bj}), \qquad \bPh^{\bar a}\to
f^*(q^a_i,\T^{\bj})\Z^{\bar a},\label{tduality}\eeq
where $q^a_i$ are the modular weights of $\Phi^a$, and,
under \ux\, transformations,
\beq\vx \to \vx + \Lambda_X + \bar\Lambda_X,\qquad
\Phi^a \to e^{-q^a_X\Lambda_X}\Phi^a,\qquad
\bar\Phi^a \to e^{-q^a_X\bL_X}\bar\Phi^a, \eeq
where $\vx$ is the \ux\, vector superfield, with $\Lambda_X\;
(\bar\Lambda_X)$ chiral (antichiral).  In the regulated theory the
anomalous part of the Lagrangian takes the form~\cite{bganom}%
\beq \L_{\rm anom} =  {1\over8\pi^2}\superint\Tr\(\eta\Omega_m\ln\cM^2\)
,\label{Lanom}\eeq
where $\cM^2$ is a real superfield whose lowest component is the PV
squared mass matrix:
\beq \l\cM^2\r = |m(z,\z,\l\vx\r)|^2,\eeq
with $z,\z,\l\vx\r$ the lowest components, respectively, of
$Z,\Z,\vx$, and $\eta = {\rm diag}(\pm1)$ is the PV signature matrix.  
Under a general anomalous transformation the logarithm in
\myref{Lanom} shifts by an amount
\beq \Del\ln\cM^2 =  H_m(T^i,\Lambda_X) + \H_m(\T^{\ibar},\bar\Lambda_X),\eeq
with $H_m$ a (matrix-valued) chiral superfield.  The resulting anomaly is given
by~\cite{bganom,dan}
\bea \Del\L_{\rm anom} &=& 
 {1\over8\pi^2}\superint\Tr\[\eta\Omega_m H_m(T,\Lambda_X)\]\hc,\label{delL}\eea
\bea\Omega_m &=& -{1\over48}\[\cM^2\bchiproj\cM^{-2}R^m\hc\] -
{1\over24}G_m^{\alpha\dot\beta}G^m_{\alpha\dot\beta} - {1\over6}R^m\bR^m
\ddd + {1\over3}{\Omega}_W
+ \Omega_{\rm YM} - {1\over36}\Omega_{X^m},\label{Om}\eea
where the operators in \myref{Om} are defined by
\bea R^m &=& -{1\over8}\cM^{-2}\chiproj\cM^2, \qquad 
 G^m_{\alpha\dot\beta} = \half\cM[\Da,\Dd]\cM^{-1}
+ G_{\alpha\dot\beta},\label{defORG}\\
\chiproj{\Omega}_W &=& W^{\alpha\beta\gamma}W_{\alpha\beta\gamma},
\qquad \chiproj\Omega_{\rm YM} = \sum_{a\ne X}T^2_a\Wc_a\Wa^a,
\label{defOJ}\\
\chiproj\Omega_X^m &=& \Xc_m\Xa^m,\qquad
\Xa^m = {3\over8}\chiproj\Da\ln\cM^2 + \Xa.\label{defOX}\\
\Xa &=& - {1\over8}\chiproj\Da K,\label{defOXm}\eea
The superfields $R$ and $G_{\alpha\dot\beta}$ are related to elements
of the super-Riemann tensor; their lowest components are auxiliary
fields of the supergravity supermultiplet. $K$ is the K\"ahler
potential, and $W_{\alpha\beta\gamma}$ and $\Wa^a$ are the superfield
strengths for, respectively, spacetime curvature and the Yang-Mills
gauge group with generators $T_a$. We are (almost) working in K\"ahler
\uk\, superspace~\cite{bggm}, where the superdeterminant of the
supervielbien $E$ is related to the superdeterminant $E_0$ of
conventional superspace by a superWeyl transformation: $E = E_0 \,
e^{-\frac{1}{3}K(Z,\bar{Z})}$, so that the Lagrangian for the
supergravity and chiral supermultiplet kinetic energy is
\beq \L_{\rm kin} = -3 \int\,
E_0 \, e^{-\frac{1}{3}K(Z,\bar{Z})} = -3 \int\,E.\label{Lkin}\eeq
In the \uk\, superspace formulation, one obtains a canonical Einstein
term with no need for further Weyl transformations on the component
fields.  The structure group of K\"ahler~U(1) geometry contains the
Lorentz, $U(1)_K$, Yang-Mills and chiral multiplet reparameterization
groups. Chiral multiplets $Z^i$ are {\it covariantly} chiral:
$\D_{\dot\alpha}Z^i = \Da\Z^{\ibar} = 0,$ where the covariant
spinorial derivatives $\Da,\D_{\dot\alpha}$ contain the \uk,
Yang-Mills, spin and reparameterization connections.  However, in
order to implement PV regularization and anomaly cancellation in the
presence of an anomalous \ux, it is necessary~\cite{gagi} to explictly
introduce the \ux\, vector field $\vx$ in the K\"ahler potential for
\ux-charged chiral matter, and the \ux\, gauge connection is not
included in $\Da,\Dd$, but instead arises from spinorial derivatives
of $\vx$.  Since the PV mass is proportional to the inverse of the PV
K\"ahler metric, the $\Wc_X\Wa^X$ term that is missing from the chiral
projection of $\Omega_{\rm Y M}$ in \myref{defOJ} is implicitly
included in $\Omega_{X^m}$. The superfield $\Omega_{X^m}$ in
\myref{defOXm} can be explicitly constructed~\cite{bganom} following
the procedure used to construct~\cite{gg} the Yang-Mills Chern-Simons
superfield $\Omega_{\rm Y M}$.

The result \myref{delL}, \myref{Om} has been obtained using both
component field~\cite{bganom} and superfield~\cite{dan} calculations.
It can be shown~\cite{bganom} that PV regulation can be done in such a
way that a) gauge and superpotential couplings that contribute to the
renormalization of the K\"ahler potential $K(Z,\Z)$, as well as all
dilaton couplings, can be regulated in a T-duality and \ux\, invariant
manner, and b) the remaining anomaly can be absorbed into the masses
of chiral PV superfields with a very simple, T-duality and \ux\,
invariant, K\"ahler metric.  Given these results, it suffices to
calculate the contribution from the latter set of PV fields to obtain
the anomaly.  The new ``D-terms'', that is, the first three terms in
\myref{Om}, as well as $\Omega_{X^m}$, can be obtained most easily in
superspace, by first working in superconformal supergravity, and then
fixing the gauge to \uk\, superspace~\cite{dan}.

Anomaly cancellation is most readily implemented using the linear multiplet 
formulation for the dilaton~\cite{linear}.  
A linear supermultiplet is a real supermultiplet that satisfies
\beq (\D^2 - 8\bar R)L = (\bar\D^2 - 8R)L = 0.\eeq
It has three components: a scalar, the dilaton $\ell = \l L\r$,
a spin-$\half$ fermion, the dilatino $\chi$, and a two-form $b_{\mu\nu}$
that is dual to the axion $\im s$, and no auxiliary field.
For the purpose of anomaly cancellation we want instead to use a real
superfield that satisfies the {\it modified} linearity condition:
\beq (\bar\D^2 - 8R)L = -\Phi,\qquad (\D^2 - 8\bar R)L = 
- \bar\Phi,\label{lincond}\eeq 
where $\Phi$ is a chiral multiplet with \uk\, and Weyl weights~\cite{bggm}
$w_K(\Phi) = 2,\; w_W(\Phi) = 1.$
Consider a theory defined by the K\"ahler potential $K$ and
the kinetic Lagrangian $\L$:
\beq K = k(L) + K(Z,\Z),\qquad  \L = -3\displaystyle{\superint}\,E\,F(Z,\Z,\vx,L)
\label{Llin1}.\eeq
When a (modified) linear superfield $L$ is included, the condition \myref{Lkin} for a
canonical Einstein term in \uk\, superspace is replaced by
\beq
F- L\frac{\partial F}{\partial L} = - L^2{\pp\over\pp L}\({1\over L}F\)
 = 1- \frac{1}{3}L \frac{\partial k}{\partial L},\label{eincond}\eeq
with the solution:
\beq F(Z,\Z,\vx,L) = 1 + \frac{1}{3} L V(Z,\Z,\vx) +
\frac{1}{3}L \int \frac{d L}{L} \frac{\partial k(L)}{\partial
L},\eeq
where $V$ is a constant of integration, and therefore independent of $L$.
If we take  
\bea V &=& - bV(Z,\Z) + \dx V_X,\\ 
V(Z,\Z) &=& \sum_i g^i + O(e^{\sum_i q^a_i g^i}|\Phi^a|^2), 
\qquad g^i = -\ln\(T^i + \T^{\ibar}\), \\ 
8\pi^2b &=& C_a - C^M_a + 2\sum_b C^b_a q^b_i + b^a_i\quad \forall\quad i,a,
\label{bconds}\\
4\pi^2\dx &=& - {1\over24}\Tr T_X = - 
{1\over3}\Tr T_X^3 = - 
\Tr(T^2_aT_X)\quad \forall\quad a\ne X\label{uxconds},
\eea
such that under an anomalous transformation $\Del V = H(T,\Lambda_X) +
\H(\T,\bar\Lambda_X)$, then
\beq \Del\L =  {1\over8}\superint{E\over R}\chiproj L H\hc =
-{1\over8}\superint{E\over R}\Phi H\hc,\label{delL2}\eeq
since the term involving $\bD^2$ vanishes identically~\cite{bggm}.
The anomaly \myref{delL} will be canceled: $\Del\L = -\Del\L_{\rm anom}$,
provided \myref{delL} reduces to the form
\bea \Del\L_{anom} &=&   - \superint\Omega H(T,\Lambda_X)\hc,\\
\Omega &=& - \Tr\[{c_d}\lbr\cM^2\bchiproj\cM^{-2}R^m\hc\rbr +
{c_g}G_m^{\alpha\dot\beta}G^m_{\alpha\dot\beta} + {c_r}R^m\bR^m\]
\ddd + {c_w}{\Omega}_W
+ \Tr\(c_{a}\Omega^a_{\rm YM} - {c_m}\Omega_{X^m}\),\label{O}\eea
where the (matrix valued) constants $c_n = \eta c'_n(q_i,q_X)$ depend
on the signatures, modular weights $q_i$ and \ux\, charges $q_X$ of
the PV fields.  They are determined by the requirement that quadratic,
linear and logarithmic divergences cancel, and will be given
explicitly in~\cite{bganom}. In particular, we require and $c_w = 8$,
$c_a =1$ in the class of models we are considering with affine level
$k_a = 1$. The resulting component expression includes the standard
chiral anomaly, including~\cite{danf} contributions from the K\"ahler
\uo\, and reparameterization connections.  In the present approach,
the factor 1/3 in the coefficient of $F^X_{\mu\nu}\tF_X^{\mu\nu}$,
relative to that of $F^a_{\mu\nu}\tF_a^{\mu\nu}$, comes from a
combination of the operators $\Omega_{X^m}$ and
$G^m_{\alpha\dot\beta}G_m^{\alpha\dot\beta}$ in \myref{defORG}.

Now consider the following Lagrangian
\beq\L_{\rm lin} = -3\superint \, E \[F(Z,\Z,\vx,L) 
+ \frac{1}{3} (L + \Omega)(S+\bar{S})\],\label{Llin}\eeq
where $S\;(\S)$ is chiral (antichiral):
\beq S = \chiproj\Sigma,\qquad \S = \bchiproj\Sigma^{\dag},\qquad
\Sigma\ne\Sigma^{\dag},\eeq
with $\Sigma$ unconstrained; $L= L^\dag$ is real but otherwise unconstrained,
and $\Omega$ is the anomaly coefficient \myref{O}:
\beq (\bar\D^2 - 8R)\Omega = \Phi,\qquad (\D^2 - 8\bar R)\Omega = \bar\Phi.\eeq
If we vary the Lagrangian \myref{Llin} with respect to the unconstrained
superfields $\Sigma,\Sigma^\dag$, we recover the modified linearity condition
\myref{lincond}. This results in the term proportional to $S+\S$ dropping out
from  \myref{Llin}, which reduces to \myref{Llin1}, with
\beq  F(Z,\Z,\vx,L) = 1 - {1\over3}\[2L s(L) - V(Z,\Z,\vx)\],\qquad
s(L) = - \half\int{d L\over L}{\pp k(L)\over\pp L},\eeq
where the vacuum value $\langle{\l s(L)\r}\rangle = \langle{s(\ell)}
\rangle = g^{-2}_s$ is the gauge coupling constant at the string scale.

Alternatively, we can vary the Lagrangian \myref{Llin} with respect to
$L$, which determines $L$ as a function of $S + \S + V$, subject
to the condition
\beq F + {1\over3}L(S + \S) = 1,\label{einst}\eeq
which assures that once the (modified) linear multiplet is eliminated, the
form \myref{Lkin}, with a canonically normalized Einstein term, is recovered.
Together with the equation of motion for $L$, the condition \myref{einst}
is equivalent to the condition \myref{eincond}, and the Lagrangian \myref{Llin}
becomes
\beq\L_{\rm lin} = -3\superint\,E - \superint\,E(S + \S)\Omega
= -3\superint\,E + {1\over8}\(\superint{E\over R}S\Phi\hc\).\label{Lchi}\eeq
Since $L = L(S + \S + V)$ is invariant under T-duality and \ux, we
require $\Del S = - H$, so the variation of \myref{Lchi} is again
given by \myref{delL2}.  The above duality transformation can be
performed only if the real superfield $\Omega$, with K\"ahler weight
$w_K(\Omega) = 0$, has Weyl weight $w_W(\Omega) = - w_W(E) = 2$, so
that $E\Omega = E_0\Omega_0$ is independent of $K$ and therefore Weyl
invariant and independent of $k(L)$.  The operator \myref{O} indeed
satisfies this requirement, as has been verified~\cite{dan} by
identifying the Weyl invariant operators in conformal superspace, and
then gauge-fixing to \uk\, superspace.

The Lagrangian \myref{Lchi} includes new tree level couplings that
generate new ultraviolet divergences.  We expect that these can be
regulated by PV fields with modular and \ux\, invariant masses, as was
shown~\cite{bganom} to be the case for the dilaton coupling to
$\Phi_{YM}$, so they will not contribute to the anomaly. These new
terms are in fact expected from superstring-derived supergravity.  The
Lagrangian depends on the 2-form $b_{\mu\nu}$ only through the 3-form
$h_{\mu\nu\rho}$.  For a linear multiplet, the 3-form is just the curl
of the 2-form:  $h_{\mu\nu\rho} = \pp_{[\mu}b_{\nu\rho]}.$
This is modified by \myref{lincond}.  In 10d supergravity we have
\beq H_{L M N} = \pp_{[L}B_{M N]}  + \omega^{\rm YM}_{M N L}
+ \omega^{\rm Lor}_{M N L}, \qquad M,N,\ldots = 0,\ldots9,\eeq
where $\omega^{\rm YM}$ and $\omega^{\rm Lor}$ are, respectively, the
10d Yang-Mills and Lorentz Chern-Simons forms.  When this theory is
compactified to 4d supergravity, we obtain the 4d counterparts of the
Yang-Mills and Lorentz Chern-Simons forms, as well as additional terms
that arise from indices $m,n,\ldots = 4,\ldots,9,$ in the compact 6d
space:
\beq h_{\mu\nu\rho} = \pp_{[\mu}b_{\nu\rho]} + \omega^{\rm YM}_{\mu\nu\rho}
+ \omega^{\rm Lor}_{\mu\nu\rho} + {\rm scalar\; derivatives} + \ldots,
\qquad \mu,\nu,\ldots = 0,\ldots,3.\eeq

To conclude, we have determined the general form of the supergravity
anomaly, and described how it may be canceled by a generalized
Green-Schwarz mechanism.  In many compactifications the anomaly is not
completely canceled by the GS mechanism and string loop threshold
corrections play a role; these are reflected in the parameters
$b_i^a$ in \myref{bconds}. They can easily be incorporated into the present
formalism by introducing~\cite{bganom} a dependence on the T-moduli
in the superpotential for the massive PV fields: $W_{P V}=\mu(T^i)
Z_{P V}Z'_{P V}$.  Phenomenological applications of our results as well
as a more precise connection to the underlying string theory will be
explored elsewhere.

\vskip .3in
\noindent{\bf Acknowledgments.} One of us (MKG) acknowledges the hospitality
of the Kavli Institute for Theoretical Physics, where part of this work 
was performed. This work was supported in part by
the Director, Office of Science, Office of High Energy and Nuclear
Physics, Division of High Energy Physics, of the U.S. Department of
Energy under Contract DE-AC02-05CH11231, in part by the National
Science Foundation under grants PHY-0457315 and PHY05-51164.

\end{document}